\documentclass[epj,nopacs]{svjour}
\usepackage{graphics}
\usepackage{graphics}
\usepackage{graphicx}
\usepackage{epsfig}
\usepackage{amssymb}
\usepackage{hyperref}
\usepackage{latexsym}
\usepackage{mathrsfs}
\usepackage{amsmath}
\usepackage{amscd}
\usepackage{xcolor}
\usepackage{verbatim}
\usepackage{soul}
\begin{document}
\title{The effect of temperature on generic stable periodic structures in the
    parameter space of dissipative relativistic standard map}
\author{Ana C.C. Horstmann \and Holokx A. Albuquerque\thanks{email: holokx.albuquerque@udesc.br} 
    \and Cesar Manchein\thanks{email: cesar.manchein@udesc.br}}
\institute{Departamento de F\'\i sica, Universidade do Estado de Santa
  Catarina, 89219-710 Joinville, Brazil}  
\date{Received: date / Revised version: date}
%
\abstract{In this work, we have characterized changes in the dynamics 
  of a two-dimensional relativistic standard map in the presence of 
dissipation and specially when it is submitted to thermal effects
modeled by a Gaussian noise reservoir. By the addition of thermal
noise in the dissipative  relativistic standard map (DRSM) it is
possible to suppress typical stable periodic structures (SPSs)
embedded in the chaotic domains of  parameter space for large enough
temperature strengths. Smaller  SPSs are first affected by thermal
effects, starting from their borders, as a  function of
temperature. To estimate the necessary temperature strength capable to
destroy those SPSs we use the largest Lyapunov exponent to obtain the
critical temperature ($T_C$) diagrams. For critical temperatures the  
chaotic behavior takes place with the suppression of periodic 
motion, although, the temperature strengths considered in this work
are not so large to convert the deterministic features of the
underlying system into a stochastic ones.
 }
\maketitle
%
\section{Introduction}
\label{introduction}
The complex behavior of a charged particle in a field of a wave
packet represents one of the fundamental problems envolving the theory 
of plasma physics and it has attracted attention for many
years~\cite{ll92}. One simple way to characterize such complex
behaviors is the study of a discrete time version of differential
equations used to model the motion of classical particles. A well
know discrete time system or map used to describe the dynamics of 
a relativistic particle in a wave packet was originally proposed in 
Ref.~\cite{ctvz89}, revisited in \cite{nih92,mmss05} and extended to
dissipative regime in \cite{cbs02}. A natural question arises from
this scenario when one intends to submit a dissipative relativistic
particle to thermal effects modeled by Gaussian noise added to the
system: what are the most common effects caused by noise in the
periodic underlying dynamics of charged particles as function of
temperature of thermal environment? 
As reported in Ref.~\cite{lt11}
there are two common phenomena induced by the influence of temperature
or noise: (i) noise-induced chaos and (ii) noise-induced crisis. In (i)
trajectory can visit the original periodic attractor (null noise) and
the chaotic saddle, creating in this way an extended chaotic
attractor; while in (ii) noise can cause trajectories on the attractor
to move out of its basin of attraction. Actually, in the real world,
classical particles always suffer environmental effects, such as noise
or temperature, which give rise to new dynamical behaviors and
therefore, requiring new comprehensive approach. Then, to explain such
behaviors in nature and for 
technological applications, it is of fundamental priority to
understand the effect of noise (thermal or not) on the resistence of
periodic behaviors.  

In the present work we revisit the investigation of nonlinear dynamics
of a particle described by a dissipative relativistic standard map 
(DRSM), introduced by \cite{cbs02} and later explored in
\cite{ly08,olr11,ol12} and, specially extend the study published by
Oliveira and Leonel in the Ref.~\cite{ol14}, where the authors
investigated the statistical and dynamical behaviors for conservative
and dissipative relativistic particles in waveguide. In order to
understand and to describe the role of dissipation and
relativistic parameters in the DRSM we perform extensive numerical
investigations besides some few analytical analysis. In addition, we
also characterize the main changes in the dynamics of such particles
under the influence of a thermal environment. Our aim is to understand 
what happens with periodic and chaotic attractors when  the
relativistic particle is affected by a Gaussian stochastic signal
for different temperature strengths. To this end we use the following
numerical approaches: (i) calculate the Lyapunov spectra (to get the
Largest Lyapunov Exponents (LLE)) based on Benettin's algorithm
\cite{bggs80,wssv85}, which includes the Gram--Schmidt
re-orthonormalization procedure and, (ii) obtain the periods of
trajectories for different pairs of parameters used to  describe
dynamical behaviors in two-dimensional diagrams (so called Lyapunov
and isoperiodic diagrams, respectively) of the four-dimensional
parameter-space of the discrete-time system. Besides,  we also
estimate the critical temperature needed  to destroy the periodic
attractors using the value of LLE.     

We find that all possible considered two-dimensional diagrams
present generic self-organized periodic structures embedded in
a single chaotic domain, namely {\it stable periodic structures}
(SPSs), well known to be typical or generic in dissipative nonlinear
dynamical systems. Such kind of SPSs were found in a wide range of
both, discrete and continuous time systems \cite{cmab11,cmab14} (and
references therein), including ratchet systems, chemical reactions,
population dynamics, electronic circuits, and laser models, among
others. In addition, as far as we know, the resistance of SPSs as a
function of temperature was recently characterized using the parameter 
planes dissipation {\it versus} ratchet parameter considering the
presence of Gaussian noise in ratchet systems as can be found in the
Ref.~\cite{mcb13}. 

In the following, we describe the organization of this paper. Starting
with Sec.~\ref{model}, we introduce the { DRSM and some
  analytical results}. Sec.~\ref{res:T0} presents the numerical
results (for null temperature case) that clearly show the influence of 
dissipation and relativistic parameters in the presence of generic
SPSs embedded in a single chaotic domain. Results about extensive
numerical experiments are discussed in Sec.~\ref{res:T} and they
confirm the destruction of SPSs (starting from their borders) and
large periodic domains when stochastic effects are considered for
large enough temperature, remaining only the chaotic ones. Finally, in  
Sec.~\ref{conc} we summarize and discuss our main conclusions.     

\section{The discrete time model}
\label{model}
Based on the well known dissipative standard map \cite{ll92} the
authors of Ref.~\cite{cbs02} introduced a dissipative
relativistic standard map (DRSM), that is a discrete time system
described as a transformation onto the real plane to itself
$f(x,y):\mathbb{R}^2\mapsto\mathbb{R}^2$, used to investigate the 
dynamics of a massive charged particle in the electric field of an
electromagnetic wave packet. The main aim was describe how 
relativistic effects change the nonlinear dynamics presented by a
dissipative standard map. The effects of the environment are taken
into account in the DRSM dynamics through of a velocity-dependent
dissipation and thermal fluctuations. In such case, the relativistic
particle is placed at a Gaussian thermal bath giving the following map
\begin{eqnarray}
  \left\{
    \begin{array}{ll}
      y_{t+1} = \gamma y_t+\dfrac{K}{2\pi}\sin(2\pi x_t)+\xi(t), \\
      \\
      x_{t+1} = x_t + {\dfrac{y_{t+1}}{\sqrt{1+(\beta y_{t+1})^2}}}
      \hspace{3mm} \mbox{mod}~1,
      \label{drsmt}
    \end{array}
  \right.
\end{eqnarray}
where $t=0,1,2,\ldots$ represents the discrete time, $y$ is the 
momentum variable conjugated to $x$, $K$ is the nonlinear parameter 
responsible for the transition from integrable motion ($K=0$) to 
chaotic dynamics ($K\neq 0$). The dissipation parameter $\gamma$ 
reaches the overdamping limit for $\gamma=0$ and the
Hamiltonian-conservative limit for $\gamma=1$ and $\beta$ represents
the relativistic parameter. In the limit of $\beta\rightarrow 0$ and
$\gamma\rightarrow 1$ the relativistic standard map is reduced to
Chirikov standard map. For $\beta=1$ the resonance case is reached and
it was previously studied in the Ref.~\cite{ly08}. The thermal noise
represented by $\xi(t)$ is a stochastic variable obeying
$\langle\xi(t)\rangle=0$ and the fluctuation-dissipation theorem that
gives the relation between $\xi$ and $\gamma$ as
$\langle\xi(t)^2\rangle=2(1-\gamma)k_BT$, with $k_B=1$ being the
Boltzmann constant and $T$ the temperature.  

\subsection{{{A brief analytical analysis}}}
\label{analytical}
Analytical boundaries for the SPSs in the two-dimensional parameter  
diagrams can be obtained for fixed points (period-1 attractors or
  orbits), as previously performed for a ratchet discrete-time system 
  in Ref.~\cite{cmab11}. Such boundaries are determined from the
analytical expression for eigenvalues $\lambda(K,\beta,\gamma,N)$
obtained from Jacobian matrix of the map (\ref{drsmt}) after one
iteration as described bellow.   

The fixed points or period-1 attractors $(x^*,y^*)$ of
Eq.~(\ref{drsmt}) for $T=0$, are obtained after one iteration by 
solving the system     
\begin{eqnarray}
  \left\{
    \begin{array}{ll}
      \sin(2\pi x^*) = \dfrac{2\pi}{K}(\gamma-1)y^* \\
      \\
      x^* = \left(x^* + \dfrac{y^*}{\sqrt{1+(\beta y^*)^2}}\right)
      \hspace{3mm} \mbox{mod}~1,
      \label{orbs}
    \end{array}
  \right.
\end{eqnarray}
which after some algebraic manipulations result in the orbital
solutions (fixed points coordinates in phase space) given by  
\begin{eqnarray}
  \left\{
    \begin{array}{ll}
      y^*_N = \dfrac{N}{\sqrt{1-(N\beta)^2}}, \\
      \\
      x^*_N = \dfrac{1}{2\pi}
      \arcsin\left[\dfrac{2\pi}{K}(\gamma-1)\dfrac{N}{\sqrt{1-(N\beta)^2}}\right],  
      \label{solorbs}
    \end{array}
  \right.
\end{eqnarray}
where $N$ is an integer or rational number which results from the
solution of second expression of Eq.~(\ref{orbs}) and, it is related
to the position of respective fixed point in phase space. As a term 
under a square root must be positive to be real and the $\vert
\sin(2\pi x^*) \vert \le 1$, we obtain the restrictions $\vert N \vert
< 1/\beta$ and $\vert (2\pi /K) (\gamma-1) N (1-(N\beta)^2)^{-1/2} \vert
\le 1$, which is the reason for a finite number of fixed points in
phase space. 

Substituting the orbital solutions (\ref{solorbs}) in the Jacobian
matrix of map (\ref{drsmt}) just after one iteration, we find the
analytical expressions for two eigenvalues $\lambda(K,\beta,\gamma,N)$
as a function of system's parameters and $N$. Setting one of these 
eigenvalues equal to $+1$ (for more details see Ref.~\cite{gh83}), we
find where (and for which parameter sets) period-1 attractors are born
in the phase-space. As the Jacobian matrix and eigenvalue
expressions are very large they are not written explicitly here. So,
using this procedure it is possible to obtain $K(\beta,\gamma,N)$
and $\beta(\gamma,K,N)$, which defines the  boundaries in the
two-dimensional parameter diagrams where period-1 attractors are
born. In  other words, the solution for equation
$\lambda(K,\beta,\gamma,N)-1=0$ for one eigenvalue results in the
following expression    
\begin{equation}
  K(\beta,\gamma,N) = \pm\frac{2\pi
    (\gamma-1)N}{\sqrt{1-N^2\beta^2}}.
  \label{kbeta}
\end{equation}
Otherwise, if the eigenvalue equation is solved for $\beta$ in terms
of $\gamma,K$ and $N$ one obtains  
\begin{equation}
  \beta(\gamma,K,N\ne 0)=\sqrt{\frac{K^2-4\pi^2(\gamma-1)^2N^2}{N^2K^2}}.
  \label{betak}
\end{equation}
In summary, these expressions are very useful for allowing us to find
position birth of the period-1 attractors in the ($K,\beta$) (or
$\beta$,$\gamma$) two-dimensional Lyapunov or isoperiodic diagrams,  
respectively, as discussed in the next section (see
Fig.~\ref{fig1}(b)). An additional detailed discussion about 
  a general procedure to find analytical expressions used to fit
  numerical boundaries between domains, characterized by different
  stable motions, in two-dimensional parameter space of generic 
  two-dimensional dynamical systems can be found in Ref.~\cite{g95}
  (see also references therein).

\section{{{Numerical experiments for $T=0$}}}
\label{res:T0}
This Section is devoted to discuss the role of relativistic parameter
$\beta$ in the dynamics of DRSM without the presence of thermal
effects, {\it i.e.}, $T=0$ in Eq.~(\ref{drsmt}). We show the results
applying numerical methods and analytical analysis. The numerical
methods are based on the iteration of the map  and obtain the
two-dimensional diagrams for the largest Lyapunov exponent (Lyapunov
diagram) \cite{ol12,mcms13,hsma14,srsab16} (see  also references
therein), and for the periods of the attractors (isoperiodic diagram)
\cite{jg12,mepzgg16,g16,harmc16} as two parameters of the studied map
are varied. 
\begin{figure*}[!htb]
  \centering
  \includegraphics*[width=0.98\linewidth]{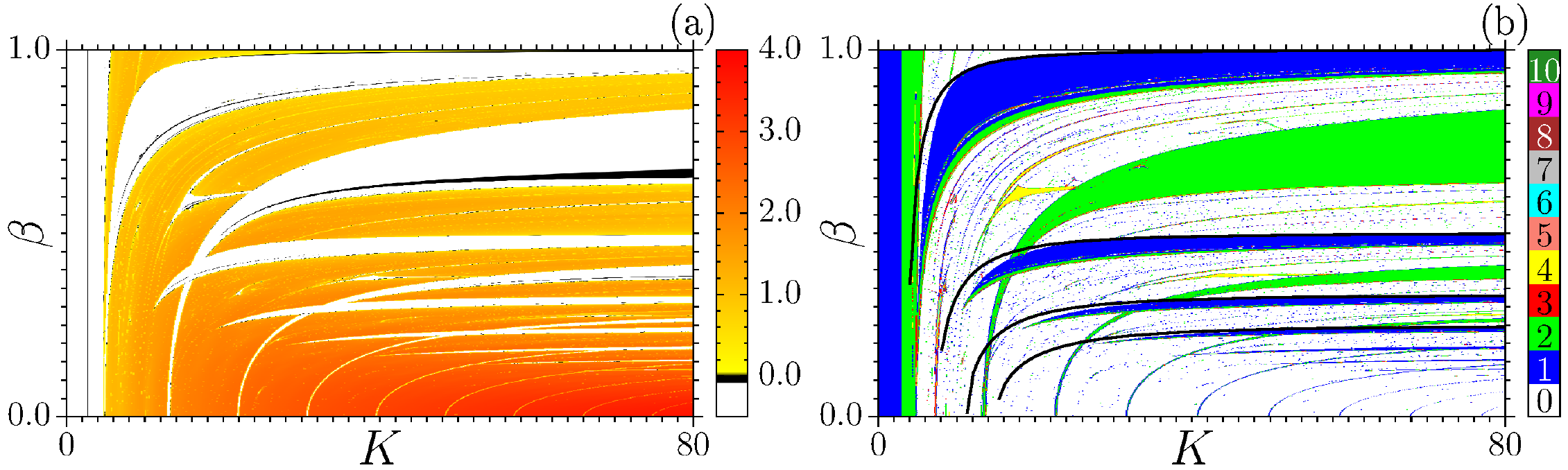}
  \caption{(a) The largest Lyapunov exponent (see color-bar) is
    plotted in the parameter-plane ($K,\beta$) for $\gamma=0.4$, while
  in (b) the periods for the same parameter-plane are plotted in
  colors (see color-bar), where the white color represents the larger
  periods than ten and chaotic attractors. The black-continuous
  curves fitting the fixed point domains (per-1 attractors) are the
  theoretical prevision given by Eq.~(\ref{betak}) for $N=1,2,3$ and
  $4$.} 
  \label{fig1}
\end{figure*}

The Lyapunov diagrams were obtained combining the set of parameters
${K, \gamma, \beta}$ taken two by two with the largest Lyapunov
exponent being evaluated as the map is iterated. The graphic
representation of the exponents is given by their magnitudes codified
by color gradients. The white represents negative exponents (fixed
points or period-1 attractors and higher-order periodic attractors),
the black null exponents (bifurcations points) and continuously
variation of yellow up to red for positive exponents (chaotic
regions). The Lyapunov diagrams were obtained considering the
discretization of the parameters in a grid of $1000 \times 1000$
values, where a typical random initial condition (IC) given by
$(x_0,y_0) =(0.1,0.3)$ was used to all parameter pairs. Some tests
using different IC sets were performed and Lyapunov diagrams remain
practically unchanged.  

On the other hand, the isoperiodic diagrams also were obtained varying
two of the set parameters ${K, \gamma, \beta}$ keeping the third
constant in a grid of $1200 \times 1200$ values of parameters, and
iterating the map with the same initial conditions listed at the
previous paragraph. In this calculations we discard $10^7$
iterations as transient time before the period of the attractor
is evaluated for  periods between $1$ up to $10$ plotted in the
isoperiodic diagrams with a discrete color scheme. For periods greater
than $10$ and  chaotic attractors, the white color is associated. At
this point becomes pertinent define an abbreviation for the
period-$q$, where $q$  represents the period of attractor, shortly
written as per-$q$, that is used along the text.  Besides the
numerical experiments we also use the analytical results discussed in 
Subsec.~\ref{analytical} to obtain the coordinates in the
two-dimensional diagrams, where the per-$1$ attractors birth as a
function of specific parameter pairs of DRSM.


In Figs.~\ref{fig1}(a)-(b), we present the Lyapunov and Isoperiodic
diagrams, respectively, for the Eq.~(\ref{drsmt}) in the ($K,\beta$)
two-dimensional diagrams. In Fig.~\ref{fig1}(a), we observe a large
chaotic region (from yellow { up} to red color) with
embedded periodic windows. In Fig.~\ref{fig1}(b), we observe that for
the white structures in (a), that could be fixed points or periodic 
attractors, now they are blue (per-1 attractors), green (per-2
attractors), red (per-3 attractors), and so on up to dark-green (per-10
attractors), as indicated by color-bar at right in the isoperiodic
diagram. As in the previous Subsection the white color represents
both attractors with periods greater than $10$ or chaotic
ones.     

Comparing both Lyapunov and isoperiodic diagrams in Fig.~\ref{fig1}(a)
and (b) respectively, for $K$ values between $0$ and $5$ we observe in
(a) a white strip that extending for $\beta$ values between $0.0$ and
$1.0$ and it corresponds to a per-1 attractor (fixed point) and a
per-2 attractor in (b) (see the blue and green colors inside that strip).
Inside it there exist a period doubling bifurcation, where
the bifurcation points are not completely visible in the Lyapunov
diagram due to the numerical precision (see the thin
  black line in that last strip in (a)). Near to $K\sim 5$ and $\sim
  12$ one can observe in (a) the birth of two large periodic
  strips (white color) that in (b) is a blue strip of per-1 and a
    green strip of per-2. From those two strips appear two distinct 
    sequences of per-1 and per-2 strips with size decreasing, as $K$ 
increases and $\beta$ decreases.    

We also plotted four curves, given by Eq.~(\ref{betak}), for $N
=1,2,3,4$ in Fig.~\ref{fig1}(b), { continuous-black} lines
bordered the first four per-1 structures (blue structures from top to
bottom). { These exemplary curves are composed by
  parameter pairs $(K,\beta)$ where the system~(\ref{drsmt}) suffers
  boundary or exterior crisis bifurcations.} It is worth to observe
that those black lines, that mark where the fixed points or per-1
structures born, do not completely fit those structures in their left
bottom side. This feature is related to the choice of the initial
conditions used to iterate the map, {\it i.e.}, the used initial 
condition gave rise to an attractor near that analytically evaluated 
by Eq.~(\ref{betak}). However, it is difficult to choose initial
conditions that exactly give rise to the same results found
analytically. { One can also see that four curves abruptly
  finish at left-hand side of same figure. This happens because
  exactly in the point where the root of Eq.~(\ref{betak}) becomes
  negative, {i.e.}, the function $\beta(K,\gamma,N)$ becomes imaginary
  and our theoretical prediction is inappropriate.} Besides, it is
possible to obtain curves for $N > 4$, but the results are not easily
visible in the present two-dimensional diagrams because the figures
resolution.   

\begin{figure}[!htb]
  \centering
  \includegraphics*[width=0.98\columnwidth]{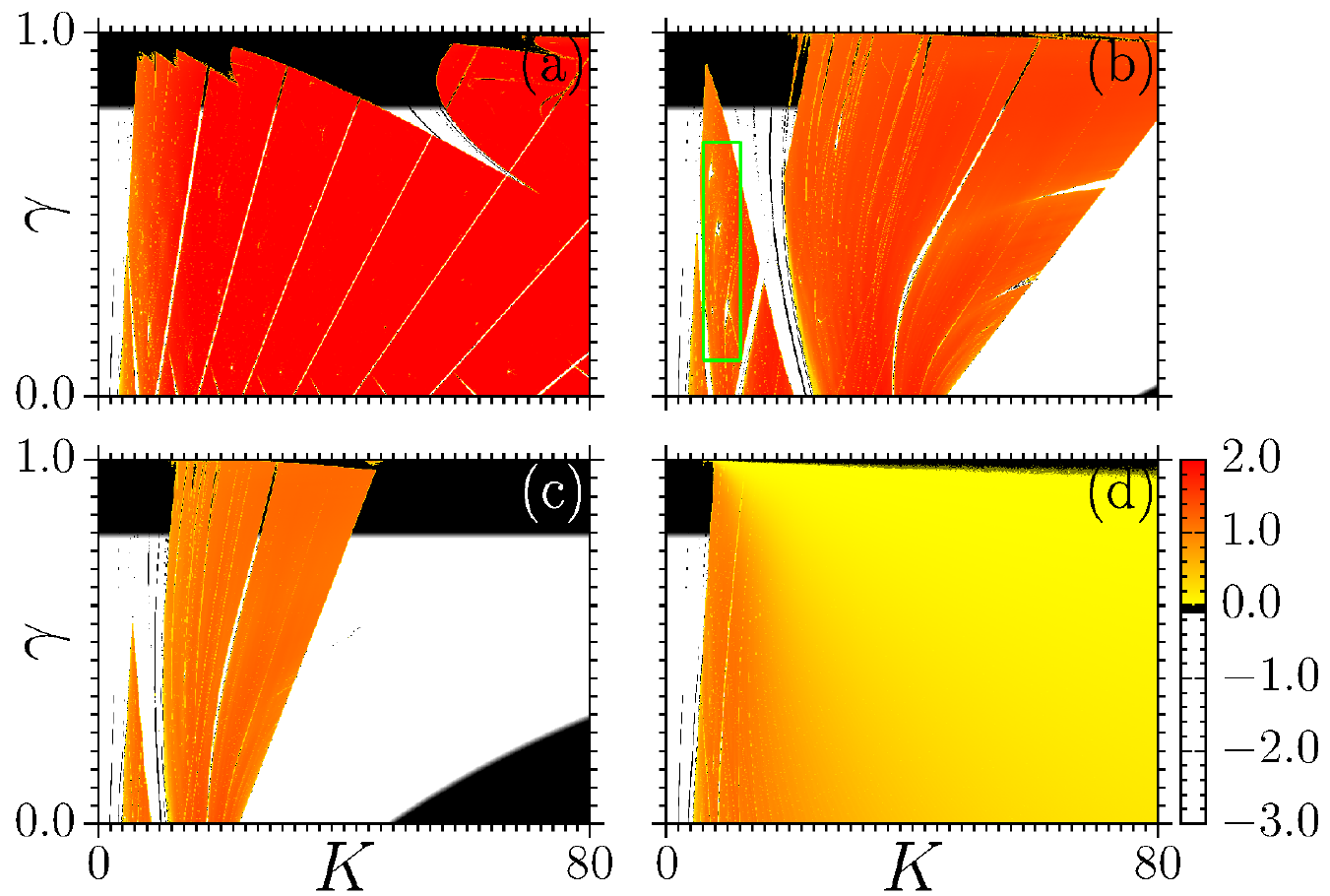}
  \caption{The largest Lyapunov exponent (see color-bar) is plotted in
    the parameter-plane ($K,\gamma$) for (a) $\beta=0.1$, (b)
    $\beta=0.4$, (c) $\beta=0.7$, and (b) $\beta=1$ (ressonant case).}
  \label{fig2}
\end{figure}

In the next, a numerical description of the DRSM is carry out,
emphasizing the relativistic parameter $\beta$. For this purpose
some Lyapunov diagrams are constructed, as described in the beginning
of Section~\ref{res:T0}. In Fig.~\ref{fig2} we show four Lyapunov
diagrams for the $(K,\gamma)$ plane, varying the parameter $\beta$. In
those four diagrams, the magnitude of the largest Lyapunov exponent 
follows the codification presented in the color-bar at right in
diagram (d). For a small $\beta$ in Fig.~\ref{fig2}(a), we observe a
periodic region (white and black color) at the left side
of the diagram and finished at the top. In addition, we also observe
the presence of some thin strips (white and black color) that cross
the chaotic region (yellow and red region) in the diagonal.
In Fig.~\ref{fig2}(b), a large periodic region (see black and
  white colors in color-bar) was born at right of the diagram,
confining the chaotic behavior into two regions. Increasing the
$\beta$ parameter, in Fig.~\ref{fig2}(c), we note the shrinkage of the
chaotic regions dominated by the periodic region at right. In the
resonant case, $\beta = 1$ in Fig.~\ref{fig2}(d), the behavior
previously observed, {\it i.e.}, the increase of the periodic region
at right in the diagrams, changes drastically, with the increase of
the chaotic domain (yellow region) destroying the periodic domain at
right of the diagram. It is also worth to note the periodic domain at
the left of the diagrams of Fig.~\ref{fig2}, that extending from
$\gamma = 0$ up to $\gamma = 1$. That region practically remains
unchanged for all values of $\beta$ used in the numerical simulations.  

\begin{figure}[!htb]
  \centering
  \includegraphics*[width=0.97\linewidth]{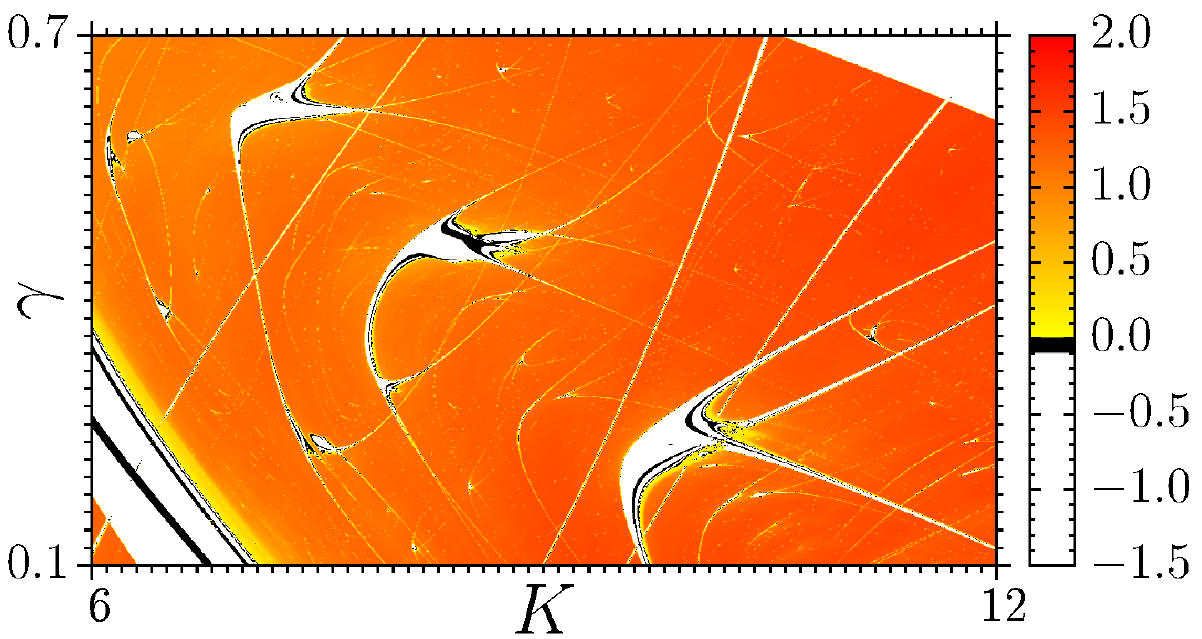}
  \caption{Magnification of Fig.~\ref{fig2}(b) highlighted by
    green box and showing typical shrimp-shaped domains.}
  \label{fig3}
\end{figure}

\begin{figure}[htb]
  \centering
  \includegraphics*[width=1.0\columnwidth]{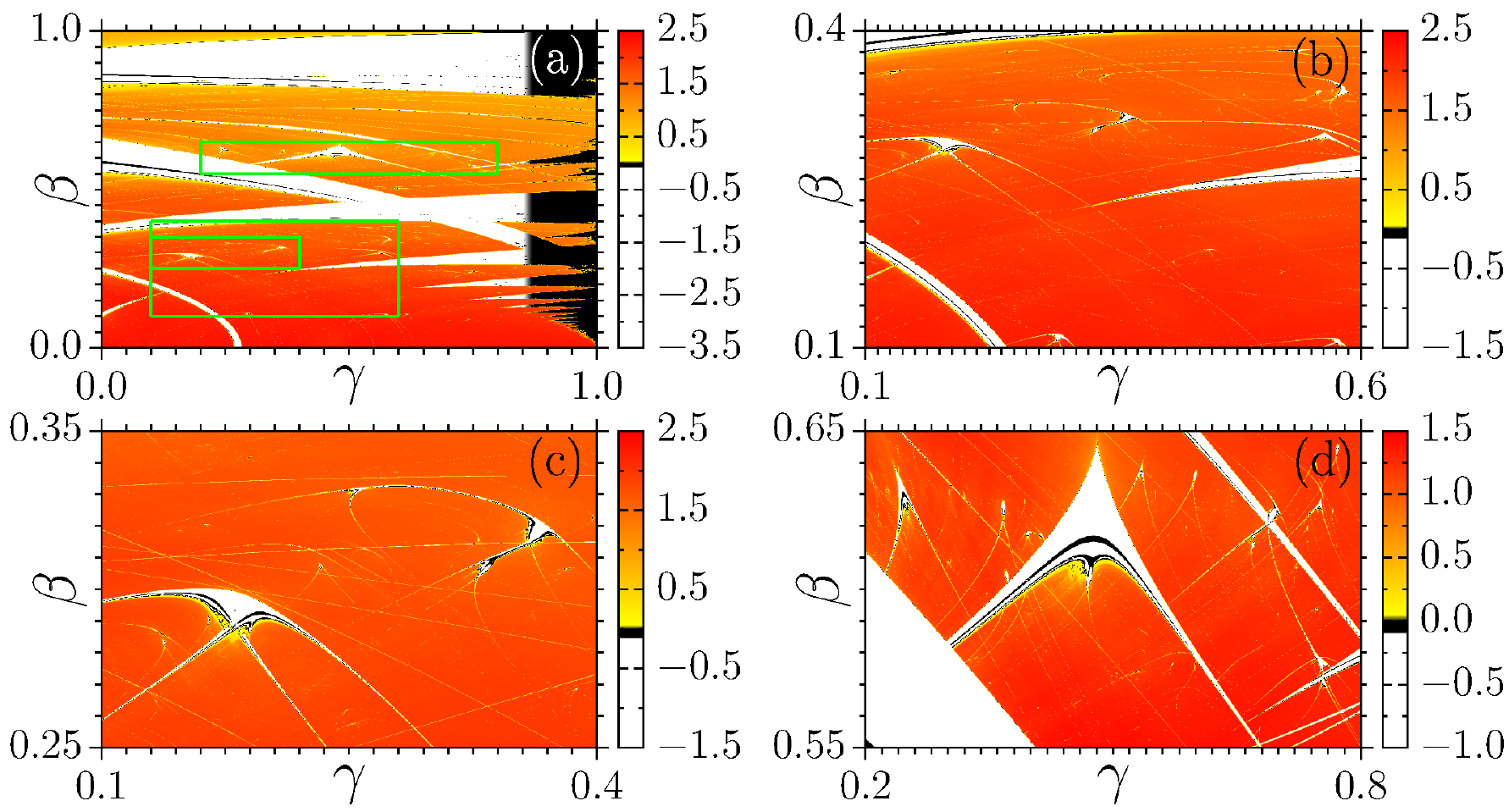}
  \caption{The largest Lyapunov exponent (see color-bar) is plotted in
    the parameter-plane ($\gamma,\beta$) for $K=20$. Figures (b-d) are
    magnifications of boxes plotted in (a).}
  \label{fig4}
\end{figure}

Inside the chaotic domains of the diagrams in Fig.~\ref{fig2}, we note
some specific SPSs in addition to the thin strips discussed above. To
clarify this, we magnified the region enclosed to the green box in
Fig.~\ref{fig2}(b), and show in Fig.~\ref{fig3}, Lyapunov diagram of
that domain: three \cite{cmab11,cmab14,fpg13} large SPSs are clearly
observed. In special, the typical SPS located in the right-bottom side
of Fig.~\ref{fig3} (about $(K,\gamma)\sim(9.5,0.25)$) is commonly
known as {\it shrimp}-shaped robust isomorphic domain (as named by
Gallas \cite{g93}) or just  {\it shrimp-}shaped domain. This kind of
SPS is formally defined by the authors of Ref.~\cite{vgg11} in
terms of superstable parabolic arcs (whose definition is given
for one-dimensional maps and recently extended for the dissipative
H\'enon map in Ref.~\cite{fog13} through of numerical continuation
methods). Another interesting definition for such generic SPSs,
named {\it compound windows} and, also characterized in parameter
space of dissipative H\'enon map as performed in \cite{g93}, was
introduced in 2008 by E. Lorenz \cite{l08}. According to Lorenz,
{\it compound windows} are periodic windows which have a typical
shape and consist of a central main ``body'' from which four narrow 
``antennae'' extend. They are organized in bands which is known as  
{\it window streets}. We have been considered such definition in
the discussion of our results, although, along the text we have
been used the expression shrimp-shaped domain. As confirmed by a
set of numerical experiments (magnifications not shown here), in some
specific portions of parameter space of DRSM there are a large number
of them. As a matter of fact, as we enlarged small portion of chaotic
regions, further smaller typical shrimp-shaped domains are found 
\cite{g93,ol11}. As far as we know, the shrimp-shaped domain also
is known as fishhook, since 1982 by Fraser and Kapral~\cite{fk82}, for
discrete maps and flows.  In addition, it has been called by different
names in the proper literature as, by instance, swallow~\cite{mh89}
and, crossroad area~\cite{cmbst91}.       

The black lines inside the shrimp-shaped domains are lines of
bifurcation, once that null largest Lyapunov exponents refer to the
bifurcations points of the map. As it was reported previously in
Ref.~\cite{hsma13} (see also references therein) the route to chaos,
from inside of a shrimp-shaped domain to the chaotic region in
specific directions, follows successive period-doubling bifurcations.   
More information about bifurcation theory and several robust results
can be found in the Ref.~\cite{m87}.   

The Lyapunov diagrams for the $(\gamma,\beta)$ plane, with $K = 20$,
are shown in Fig.~\ref{fig4}, where in (a) a large view is plotted
while in (b)-(d) magnifications are presented for regions delimited by
greens boxes in (a). In Fig.~\ref{fig4}(a) large periodic strips
(black-white strips) are seen in the diagram, and those
strips extend along the chaotic region. Such strips
represent fixed points (per-1 attractors) for each pair of
$(\gamma,\beta)$ parameters represented by white
color while black color is associated to bifurcation points (null
largest Lyapunov exponent). In addition, as showed in
Fig.~\ref{fig1}(b), the upper boundaries of those strips are
analytically fitted by Eq.~(\ref{betak}), with  $K$ fixed. In
Fig.~\ref{fig4}(b)-(d), three emblematic types of isomorphic periodic
structures embedded in the chaotic region, as reported in some recent
works about maps \cite{cmab11,g93,ol11} and flows \cite{hsma14,hsma13}
{ and discussed in the previous   paragraph} are
observed. Two of them, observed in Fig.~\ref{fig4}(b), and better
visualized in Fig.~\ref{fig4}(c), are the well-known {\it
  shrimp}-shaped domains, at left-bottom side of the diagram, while at
the right-top side of the diagram, a smaller set of three connected
SPSs. Similar structures were recently reported in
Ref.~\cite{fpg13} for flows. The third one, shown in
Fig.~\ref{fig4}(d), is the {\it cuspidal} SPSs \cite{cmab14,g95},
{ sometimes also known as saddle area~\cite{bst98} and
  rigorously demonstrated since 1985 as published in ~\cite{g91}. A
  more comprehensive and didactical explanation about cuspidal SPSs
  can be found on appendix of Ref.~\cite{gsv13}}. Those Lyapunov
diagrams also present self-affinity features, {\it i.e.}, those SPSs
appear in smaller scales of the Lyapunov diagrams.     

The second set of partial conclusions is based on the rich and complex 
dynamics presented by DRSM, even without the presence of thermal
effects, when two control parameter are varied
simultaneously. Therefore, the high-complex configuration observed in
the two-dimensional diagrams as discussed in this section is
possible due to the features of self-similarities and affinities
besides the evident proofs of existence of typical SPSs as,
shrimp-shaped and cuspidal domains, which are believed generic in the
parameter-spaces of nonlinear dynamical systems.  

\section{Using thermal effects ($T>0$) to perturb periodic attractors} 
\label{res:T}
In this Section the thermal effects are studied when the DRSM model is
added by a Gaussian noise, as described by Eq.~(\ref{drsmt}). Lyapunov
diagrams were constructed for the $( K, \gamma)$ and
$(\gamma,\beta)$ planes with $T$. 
\begin{figure*}[!htb]
  \centering
  \includegraphics*[width=1.0\linewidth]{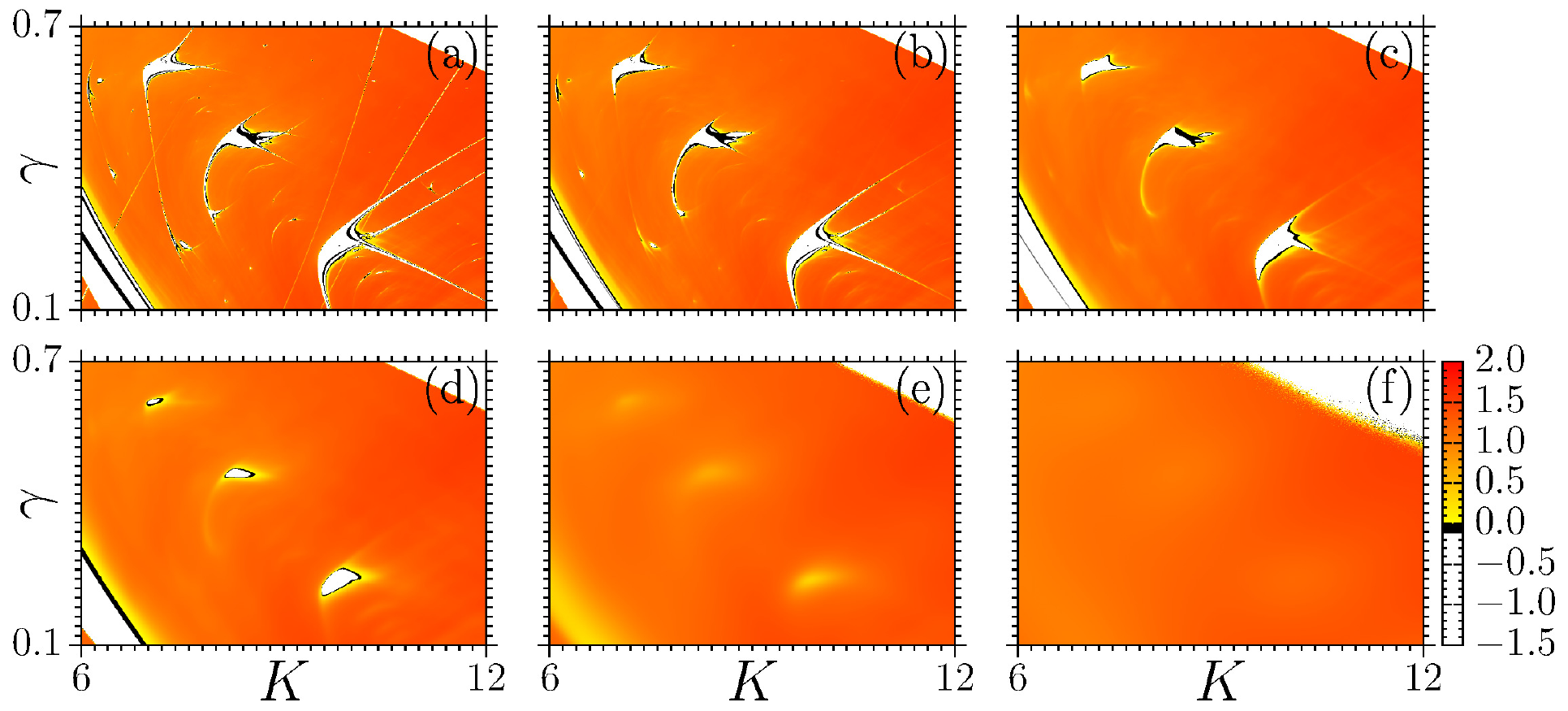}
  \caption{The largest Lyapunov exponent (see color-bar) is plotted in
    the parameter-plane $(K,\gamma)$ with $\beta=0.4$ for (a)
    $T=10^{-7}$, (b) $T=10^{-6}$, (c) $T=10^{-5}$, (d) $T=10^{-4}$,
    (e) $T=10^{-2}$, and (f) $T=10^{-1}$. As temperature increases the
    destruction of SPSs becomes more and more evident.}
  \label{fig5}
\end{figure*}
In Fig.~\ref{fig5}, the Lyapunov diagrams as $T$ increases
from $10^{-7}$ in (a), up to $10^{-1}$ in (f), are shown for 
$(K,\gamma)$. Fig.~\ref{fig3} presents the $T=0$ case. Comparing
both cases, $T = 0$ (Fig.~\ref{fig3}) with $T=10^{-7}$
(Fig.~\ref{fig5}(a)), considered very small temperature, 
minimal changes are observed. Mainly, in the chaotic region, very
small periodic structures were destroyed, and some branches of the
bigger SPSs became thin. More increasing the
temperature, Figs.~\ref{fig5}(b)-(d), this effect becomes more clear,
all smaller SPSs embedded in the chaotic region are
destroyed, just surviving the three largest, but the temperature
effect also destroyed their branches, surviving only the core of the
structures. In Figs.~\ref{fig5}(e)-(f), the temperature achieves a
critical condition where the periodic structures were completely
destroyed. It is worth to note the {\it ``scars''} that the periodic
structures leave in the chaotic region when the temperature effects
increase. For example, see in Figs.~\ref{fig5}(c)-(e), the yellowish
regions where before there were branches and pieces of periodic
structures for low values of temperatures
(Figs.~\ref{fig5}(a)-(b)). Those { pronounced} {\it scars}
{ (yellow scars in Fig.~\ref{fig5}(e))} are a kind of
``memory'' in the dynamics of the map due to the existence of the
periodic  structures. These ``memories'' can be completely
destroyed for higher values of temperatures (see Figs.~\ref{fig5}(f)).  
\begin{figure*}[!htb]
  \centering
  \includegraphics*[width=1.0\linewidth]{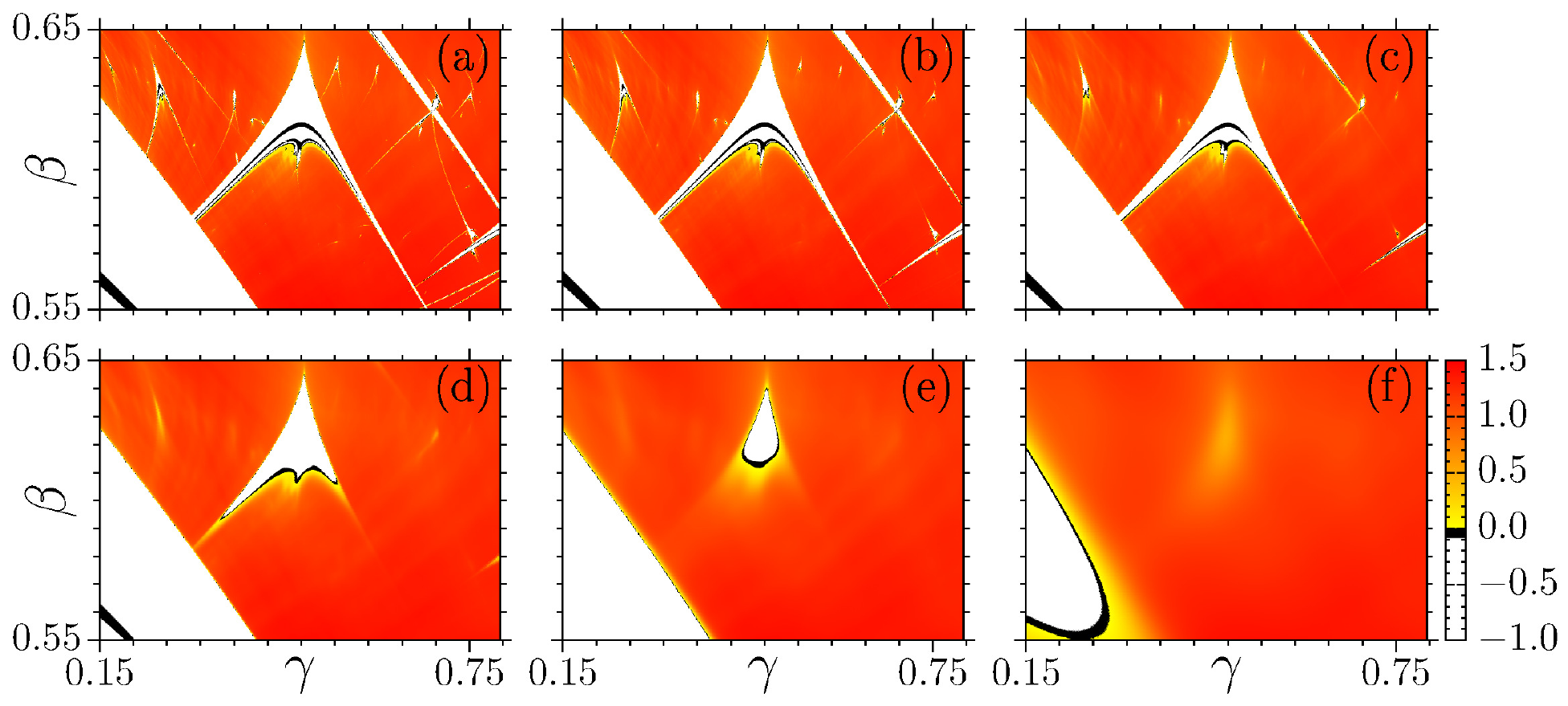}
  \caption{The largest Lyapunov exponent (see color-bar) is plotted in
    the parameter-plane $(\gamma,\beta)$ with $K=20$ for (a)
    $T=10^{-7}$, (b) $T=10^{-6}$, (c) $T=10^{-5}$, (d) $T=10^{-4}$,
    (e) $T=10^{-3}$, and (f) $T=10^{-2}$. As temperature increases the
    destruction of SPSs becomes more and more evident.}
  \label{fig6}
\end{figure*}

\begin{figure*}[!htb]
  \centering
  \includegraphics*[width=0.87\linewidth]{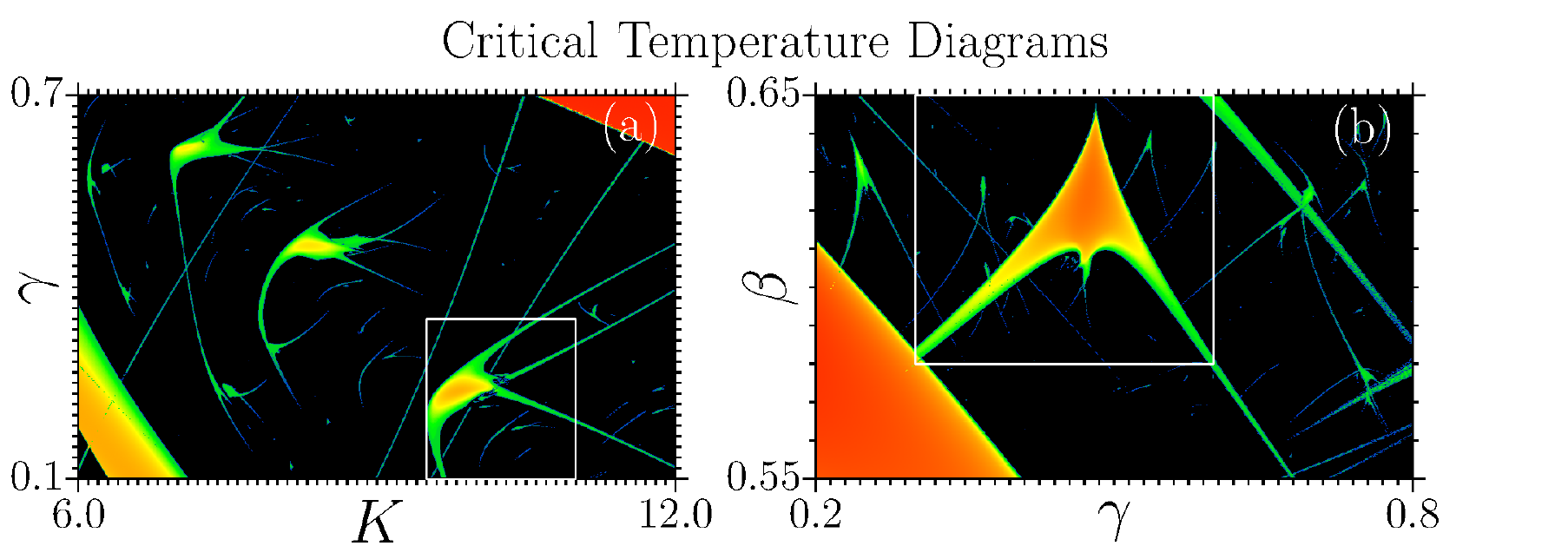}
  \includegraphics*[width=0.90\linewidth]{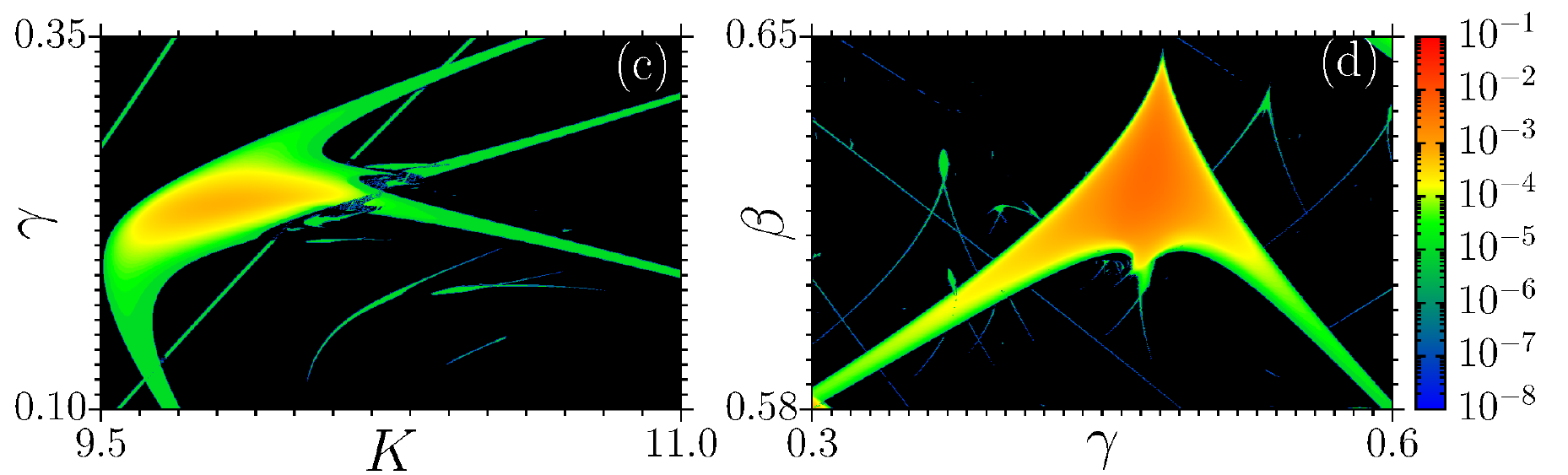}
  \caption{Plot of critical temperature $T_C$ (see the color-bar in
    (d)) necessary to destroy the SPSs in the (a) $(K,\gamma)$ plane,
    (b) $(\gamma,\beta)$ plane, and in the (c) and (d), magnifications  
    of boxes in (a) and (b), respectively.}
  \label{fig7}
\end{figure*}

In Fig.~\ref{fig6}, the Lyapunov diagrams, as $T$ increases
from $10^{-7}$ in (a) up to $10^{-2}$ in (f), are shown for
$(\gamma, \beta)$. The same behaviors presented previously in
Fig.~\ref{fig5}, are also observed in Fig.~\ref{fig6}. The size of the
structures in parameter space has a fundamental role in their
stabilities regarding the temperature effects. { Although the entire
parameter space is perturbed by considering the presence of thermal
effects in the system} the smallest SPSs are more sensitive to the
temperature { strength}, and they are the first to be
destroyed { if compared to the larger ones.}   

In the next we follow using the LLE to obtain the critical
temperatures $T_C$ of the entire structure for which it will
survive. For temperatures higher than its $T_C$, the whole
periodic structures will be destroyed by thermal effects. 
In Fig.~\ref{fig7}, we show the critical temperature diagrams for the
$(K, \gamma)$ and $(\gamma, \beta)$ planes. $T_C$ is codified by the
color-bar in a logarithmic scale. The diagrams of Figs.~\ref{fig7}(a)
and (b) are the same regions of Figs.~\ref{fig5}(a)
and.~\ref{fig6}(a), respectively. Figures~\ref{fig7}(c) and (d) are
magnifications of the boxes in (a) and (b), respectively. It is
clearly observed that $T_C$ is larger inside the structures,
especially inside their inner parts. Thus, the periodic structures can
still be recognized in such $T_C$ analysis. Even though $T_C$ is
larger inside the inner parts of the structures, it tends to decrease
at their borders. The procedure to obtain $T_C$ follows: for each pair
of parameters, $(K, \gamma)$ or $(\gamma, \beta)$, $T$ in
Eq.~(\ref{drsmt}) was increased until the LLE reached positive values
within the precision $10^{-3}$, which is a value below the minimum LLE
allowed by the periodic constraints in the structures. Other precision
values were tested and the main findings remain unchanged. It is
possible to observe that the far from zero LLE in the chaotic region
is independent of the parameters $(K, \gamma)$, or $(\gamma, \beta)$,
within the precision $10^{-3}$. This defines the critical temperature
$T_C$, where the near to zero LLE, due to periodic motion (inside the
periodic structures), is transformed into the far from zero LLE due to
the chaotic motion. By the $T_C$ analysis, we corroborate the
assumption that how bigger is the SPS in the parameter-space, more
resistant to destruction by thermal noise is the SPS.   

We finish this Section stating the following conclusion: In the
presence of a thermal reservoir, modeled by Gaussian noise, generic
SPSs (as discussed above) usually are destroyed, starting from their
borders, for large enough temperature and, characterizing this
behavior as {\it generic} in parameter space. In addition,
our results confirm that smaller SPSs are {\it first}
affected by small thermal effects ($T\rightarrow0$). 

\section{Summary and conclusions}
\label{conc}
The dissipative relativist standard map (DRSM) with Gaussian noise,
Eq.~(\ref{drsmt}), was numerically studied in this work. The
Gaussian noise represents the thermal effect in the dynamics of
the DRSM model. We start  the numerical investigations of 
DRSM without the presence of noise obtaining the Lyapunov and
isoperiodic diagrams for the dissipation parameter $\gamma$ and 
relativistic parameter $\beta$ besides the kicking strength $K$
and in this way we extend the previous investigation performed in
Ref.~\cite{ol14}. The birth of period-1 structures in the $(K,\beta)$
plane was also analytically obtained and numerically corroborated in
the Lyapunov and isoperiodic diagrams. All the parameter-planes of the
DRSM model present isomorphism regarding the presence of generic
periodic structures embedded in the chaotic domains, also
known as {\it shrimp-shaped} stable periodic structures (SPSs).

In a recent work \cite{mcb13} the thermal effect was studied in a
discrete and continuous ratchet models and the critical temperature
$T_C$ necessary to destroy the SPSs was obtained by the computation of
the classical ratchet currents in the parameter-spaces. The authors
showed that the SPSs are structures with optimal ratchet currents and
how bigger these structures are in the parameter-space, more resistant
to be destroyed by thermal effects they are. There is also the
Ref.~\cite{bsmcps15}, where the authors present results about quantum
ratchet currents that inside the SPSs have been shown to  be resistant
to reasonable temperatures and to vacuum fluctuations. For some
specific parameter combinations however, temperature and vacuum
fluctuations induced ratchet currents. Now here, our numerical results
clearly show that when a typical dynamical system is perturbed by
temperature effects, SPSs immersed in chaotic domains usually are
destroyed starting from their borders (antennae). To estimate the noise
strength necessary to destroy SPSs, we propose a new numerical method
to obtain the critical temperature $T_C$ by the computation of the
largest Lyapunov exponents in the parameter-planes of the DRSM model,
Eq.~(\ref{drsmt}). It is very important to emphasize the power of this  
technique that can be applied to obtain $T_C$ using the largest
Lyapunov exponent in any system models in contact with thermal
reservoirs, where chaotic and periodic behaviors coexist for different
parameter sets. To corroborate with this discussion, the information
about $T_C$ is very relevant to show how robust are the SPSs in the
presence of external perturbations as thermal effects. In the limit of
sufficient small temperatures one can also applies this procedure to
time series and estimate the strength of noise capable 
to destroy periodic behaviors in real experiments. In this case the
chaotic characteristics of a deterministic dynamical system can be
preserved as is often in experimental situations.  

\section*{Acknowledgments}
\begin{acknowledgement}
The authors thank CAPES, FAPESC, and C.M. thanks CNPq (all Brazilian
agencies), for financial support. { We also thank the
  Referees for the constructive remarks that improved the quality of
  the work.}
\end{acknowledgement}

{
\section*{Author contributions}
H.A.A. and C.M. conceived the simulations. A.C.C.H. and C.M. performed
the simulations and all authors discussed the results. H.A.A. and
C.M. wrote the manuscript. }

%


\end{document}